\documentclass[]{spie}  
\usepackage{amssymb,amsmath}
\usepackage[]{graphicx}

\title{Ultra-high Q whispering-gallery-mode bottle microresonators: properties and applications}

\author{Danny O'Shea,\supit{a} Christian Junge,\supit{a} Sebastian Nickel,\supit{b} Michael P{\"{o}}llinger,\supit{b} and Arno~Rauschenbeutel\supit{a}
\skiplinehalf
\supit{a}Vienna Center for Quantum Science and Technology, Atominstitut, TU Wien, 1020 Wien, Austria \\
\supit{b}Institut f{\"{u}}r Physik, Johannes Gutenberg-Universit{\"{a}}t, 55099 Mainz, Germany
}

\authorinfo{Further author information: Send correspondence to A.R.\\A.R.: E-mail: Arno.Rauschenbeutel@ati.ac.at, Telephone: +43 58801 141761}

\begin{document}
  \maketitle

\begin{abstract}
Highly prolate-shaped whispering-gallery-mode ``bottle microresonators'' have recently attracted considerable attention due to their advantageous properties. We experimentally show that such resonators offer ultra-high quality factors, microscopic mode volumes, and near lossless in- and out-coupling of light using ultra-thin optical fibers. Additionally, bottle microresonators have a simple and customizable mode structure. This enables full tunability using mechanical strain and simultaneous coupling of two ultra-thin coupling fibers in an add-drop configuration. We present two applications based on these characteristics: In a cavity quantum electrodynamics experiment, we actively stabilize the frequency of the bottle microresonator to an atomic transition and operate it in an ultra-high vacuum environment in order to couple single laser-cooled atoms to the resonator mode. In a second experiment, we show that the bottle microresonator can be used as a low-loss, narrow-band add-drop filter. Using the Kerr effect of the silica resonator material, we furthermore demonstrate that this device can be used for single-wavelength all-optical signal processing.
\end{abstract}

\keywords{Optical microresonators, Pound--Drever--Hall frequency stabilization, Kerr effect, bistability, all-optical switching}

\section{INTRODUCTION}
\label{sec:intro}
Optical microresonators have proven to be a powerful tool in studies requiring strong light-matter interaction \cite{Vah03Opt}. The light inside such resonators can be strongly confined spatially and stored for an extended period of time. Consequently, very high intracavity intensities are obtained with only moderate optical powers coupled into the resonator. This makes optical microresonators ideal tools for efficiently coupling light and matter. More quantitatively, these light-confining properties are characterized by the resonator's mode volume $V$ and its quality factor $Q$. For a given in-coupled power, the resulting intra-cavity intensity is then proportional to the ratio $Q/V$. The highest values of $Q/V$ to date have been reached with whispering-gallery-mode (WGM) microresonators \cite{Kip04Dem}. Due to their strong intensity enhancement, WGM microresonators have been successfully employed for numerous applications, ranging from microlasers \cite{San96Ver,Kli00Ver,Cai00Fib} to cavity quantum electrodynamics (CQED) \cite{Ver98Cav,Aok06Obs} and nonlinear optics \cite{Car07Vis,Hay07Opt} where they greatly enhance light-matter and  light-light interactions, respectively. WGM microresonators are monolithic dielectric structures in which the light is guided near the surface by continuous total internal reflection \cite{Mat06Opt}. Light is coupled in and out of these structures by frustrated total internal reflection. When using ultra-thin optical fibers for this purpose, the coupling can be performed with near 100~\% efficiency \cite{Spi03,Poe10}, thereby exceeding the coupling efficiencies of all other types of optical microresonators.

\begin{figure}
    \begin{center}
    \begin{tabular}{c}
    \includegraphics[width=0.75\textwidth]{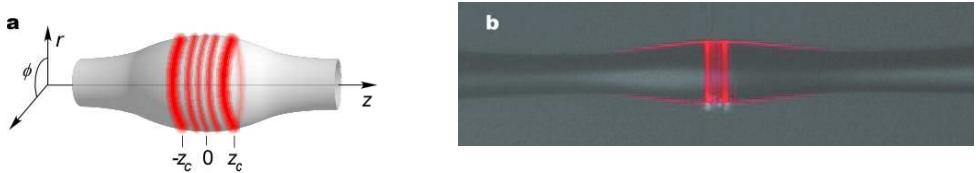}
    \end{tabular}
    \end{center}
    \caption{\label{schematic bottle resonator}
    Sketch of the resonator geometry along (a) and in a plane perpendicular (b) to the resonator axis. The Helmholtz equation is solved in cylindrical coordinates. The radial coordinate is denoted by $r$, the axial coordinate by $z$, and the azimuthal coordinate by $\phi$. A slowly varying parabolic profile of the resonator radius along the z axis is assumed (cf. Eq.~(\ref{radius profile})). Each bottle mode is located between two so-called ``caustics'' at $\pm z_c$. (b) (false color image) A micrograph of a fluorescing resonator doped with Er$^{3+}$ ions shows the actual profile of the resonator and a bottle mode.}
\end{figure}

Recently, we demonstrated a novel type of WGM microresonator with a ``bottle-like'' geometry \cite{Sum04,Lou05,Poe09} that provides a customizable mode structure while maintaining a high $Q/V$ ratio. The highly prolate shape is responsible for a class of WGMs, termed ``bottle modes'', where light oscillates back and forth along the resonator axis between two turning points which are defined by an angular momentum barrier. We fabricate these resonators from standard optical glass fibers using a heat-and-pull process. Details on the resonator fabrication can be found in Ref.~\citenum{Poe09}. As shown in Fig.~\ref{schematic bottle resonator}~(a), the resulting axial mode structure features two pronounced regions of enhanced intensity at the so-called ``caustics'', located at the turning points of the harmonic motion. While being conceptually similar to traditional WGM resonators and sharing many properties, bottle resonators have the additional advantage of combining the tunability typical to Fabry--P\'erot resonators. Thus, the resonator can be made resonant with an arbitrary frequency, fixed, for example, by an atomic transition.

The ability to tune the resonance frequency of microresonators to other frequency-critical elements is especially important when developing sophisticated optical processing applications.  For example, many CQED quantum information protocols rely on mapping the quantum state of one two-level system onto an optical field and then transferring it to another remote two-level system. Such CQED quantum networks require multiple resonators to be mutually resonant and to be linked, e.g., by optical fibers \cite{Kim08}. In these situations, it is important to be able to not only tune the resonator but to also stabilize the resonance frequency of the resonator to an external reference with a high degree of precision so as to maintain a robust system for extended periods of operation. The need for frequency stabilization in the case of ultra-high $Q$ WGM microresonators is made even more apparent by considering the typical situation where a temperature change of only 1~mK is enough to change the resonance frequency by one linewidth. A scheme used to stabilize an ultra-high $Q$ ($Q>10^8$) bottle resonator to an atomic transition using the Pound--Drever--Hall technique while in a vacuum environment is described in this article.

\begin{figure}
    \begin{center}
    \begin{tabular}{c}
    \includegraphics[width=0.7\textwidth]{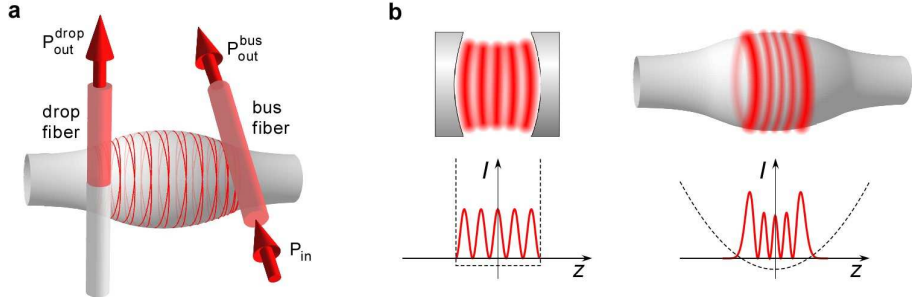}
    \end{tabular}
    \end{center}
    \caption{\label{fig_add_drop_configuration}
    (a) Schematic of the coupling setup in add-drop configuration. Two independent positioning systems are used to place two ultra-thin fiber couplers at both caustics of the bottle microresonator. Depending on its frequency, light propagating in the bus fiber is selectively coupled into the resonator mode and exits the resonator through a second ultra-thin fiber, referred to as the drop fiber. (b) Analogy between a bottle microresonator and Fabry-P\'erot microresonator. The two mirrors of a Fabry-P\'erot cavity reflect the light back and forth along the axial direction. The resulting sinusoidal intensity distribution can be understood as being due to the confinement of the photons by an effective box-like potential (dashed line). In the case of the bottle microresonator the axial confinement is caused by an effective harmonic potential fixed by the curvature of the resonator profile. The resulting intensity distribution is therefore given by the eigenfunctions of the quantum mechanical harmonic oscillator \cite{Lou05}. The intensity is significantly enhanced at the caustics of the bottle mode, located approximately at the classical turning points of the harmonic motion.
}
\end{figure}

Besides the issue of tunability, the mode geometry allows one to simultaneously access the resonator with two coupling fibers, shown in Fig.~\ref{fig_add_drop_configuration}, without the spatial constraints inherent to equatorial WGMs typically employed in microspheres and microtoroidal resonators. This facilitates the use of the bottle microresonator as a four-port device in a so-called ``add-drop configuration'' \cite{Poe10}. In communication technology, such devices are used for de-multiplexing optical signals. By judicious placement of the coupling fibers at the resonator caustics, the power transfer efficiency between the bus fiber and the drop fiber is as high as 93~\% while, simultaneously, the filter bandwidth is as narrow as 49~MHz \cite{Poe10}. This bodes well for applications requiring high efficiency.

The article is organized as follows: the spectral properties of bottle modes are briefly described in section~\ref{sec:Spectral}. The tunability and frequency stabilization of ultra-high quality  bottle resonators is presented in section~\ref{sec:Tunability} and section~\ref{sec:Stable}, respectively. Finally, in section~\ref{sec:Switch}, an application is demonstrated where a bottle microresonator is used to efficiently switch light between two coupling fibers using self-phase modulation via the Kerr effect.

\section{SPECTRAL PROPERTIES OF BOTTLE MICRORESONATORS}
\label{sec:Spectral}
The spectral properties of WGM microresonators are typically determined by their geometry. Traditional WGM resonators, like the microtoroid, microdisk, and microsphere,\cite{Vah03Opt} are generally considered 2-D resonators that confine light in an equatorial plane with their spectral properties defined by their diameter. In contrast, as shown in Fig.~\ref{schematic bottle resonator}, bottle resonators are intrinsically 3-D in design and their spectral properties are defined by their diameter as well as their curvature along the fiber axis \cite{Lou05}. The resonator radius $R$ along the resonator axis $z$ is assumed to have a parabolic profile
\begin{equation}
\label{radius profile} R(z)= R_0\cdot\left(1-1/2\left(\Delta k\cdot z\right)^2\right).
\end{equation}
Here, $R_0$ is the maximum radius of the resonator at the position $z = 0$ and $\Delta k$ denotes the curvature of the resonator profile. The electric field of a bottle mode must fulfill the Helmholtz equation
\begin{equation}
\label{} \left(\nabla^2 +  k^2\right)\cdot \vec{E} = 0~,
\end{equation}
where $\nabla^2$ is the Laplace operator, $k=\omega\cdot n/c$ is the wave number of the electromagnetic field with angular frequency $\omega$, $n$ is the refractive index of the resonator material, and the speed of light in vacuum is denoted by $c$.
Due to the prolate shape of the bottle resonator there is only a small variation in its radius along $z$, meaning that $dR/dz~\ll~1$ in the central region of the structure. This motivates the so-called ``adiabatic approximation'', meaning that the radial component $k_r = \left(dR/dz \right)\cdot k_z$ of the wave vector can be neglected with respect to the axial and azimuthal components $k_z$ and $k_{\phi}$
\begin{equation}
\label{Eqn_k}  k \approx \sqrt{k_z^2 + k_\phi^2}~ \approx \frac{2\pi n}{\lambda},
\end{equation}
where $\lambda$ is the wavelength of light in vacuum. The resonators used in this work have a diameter $D_0 = 2R_0$ of around 35--50~$\mu$m and a typical curvature of $\Delta k = 0.012$~$\mu$m$^{-1}$. The bottle modes experimentally investigated are typically located in a region $\lvert z \rvert \leq 10~\mu$m. With these values one finds $\lvert k_r(z)\rvert \leq 2.5\times 10^{-2 } \lvert k_z(z) \rvert $.

In the adiabatic approximation, the wave function is written as a product of the ``axial wave function'' $Z(z)$ and the ``radial wave function'' $\vec{\Phi}(r,R(z))$. The latter only exhibits a weak $z$ dependency via the adiabatic variation of the resonator radius. The wave equation thus reads as
\begin{equation}
\label{} \left(\nabla^2 + k^2\right)\cdot \vec{\Phi}\left(r,R(z)\right)Z\left(z\right)e^{i m \phi}=0,
\end{equation}
where the exponential component derives from the cylindrical symmetry of the azimuthal part. The radial wave equation for the component $i$ of the wave function is given by
\begin{equation}
\label{Eqn_radial_weqn} \partial_r^2\Phi_i\left(r,R(z)\right) + \frac{1}{r} \partial_r\Phi_i\left(r,R(z)\right)
+\left(k_{\phi}^2R(z)-\frac{m^2}{r^2}\right)\Phi_i = 0.
\end{equation}
It depends on the radial coordinate and contains $R(z)$ and $m$ as parameters, where $m$ is the azimuthal quantum number. It has the form of a Bessel equation, well known from the description of light propagation in optical fibers\cite{Yar91}. The radial wave equation can be solved exactly and has two linearly independent solutions; the Bessel function of the first kind $J_n(x)$ and the Bessel function of the second kind $Y_n(x)$. The general solution for $\Phi_i(r,z)$ is given by the following linear combination
\begin{equation}
\label{gen_sol_red_weq}
\Phi_i(r,z)=A \cdot J_m\left(k_{\phi}(z)\cdot r\right) + B\cdot Y_m\left(k_{\phi}(z)\cdot r\right).
\end{equation}
Additionally, WGM resonators support modes of two orthogonal polarizations; transverse magnetic modes (TM modes) where the electric field is approximately parallel to the resonator axis, and perpendicular transverse electric modes (TE modes).

The axial wave equation only depends on the $z$ coordinate
\begin{equation}
\label{Eqn_axial_weqn} \left(\partial_z^2 + k_z^2\right)\cdot Z(z) = 0.
\end{equation}
Using Eq.~(\ref{Eqn_k}) and eliminating $k_{\rm \phi}$ via $k_{\rm \phi}(z)= k \cdot R_c/R(z)= m/R(z)$, where $R_c$ is the radius at the caustic, one can solve the axial wave equation. For the parabolic radius profile of Eq.~(\ref{radius profile}), the axial wave equation is well approximated by
\begin{equation}
\label{Eqn_axial_weqn2} \partial_z^2 Z + \left(k^2 - \left(\frac{m}{R_0}\right)^2-\left(\frac{m\Delta k }{R_0}\right)^2\cdot z^2\right)\cdot Z = 0.
\end{equation}
This differential equation is equivalent to the harmonic oscillator $\partial_z^2 Z + \left(E-V(z)\right)\cdot Z = 0$~, with the ``kinetic energy'' $E$ and the ``potential energy'' $V(z)$~ as discussed in Ref.~\citenum{Lou05}.
The axial quantum number $q\in \mathbb{N}$ (nonnegative integer) gives the number of nodes in the axial intensity distribution.

The resonance condition dictates that the optical path length has to be an integer multiple of the wavelength of the light field coupled into the resonator, which is satisfied only for certain caustic radii, $R_c$. The mode spectrum is determined by the allowed eigenvalues for the wave number and is given by
\begin{equation}
\label{Eigenwerte}
 k_{m,q} = \frac{m}{R_c} = \sqrt{\frac{m^2}{R_0^2}+\left(q+1/2\right)\frac{2m\Delta k}{R_0}},
\end{equation}
with $\Delta E_m = 2m \Delta k/R_0$. The solutions for $Z$ are given by a combination of a Hermite Polynomial, $H_q$, and a Gaussian
\begin{equation}
\label{Eqn_axial_weqn5} Z_{m,q}(z) =  H_q\left(\sqrt{\frac{\Delta E_m}{2}}\cdot z\right)\exp\left(-\frac{\Delta E_m}{4}z^2\right).
\end{equation}

It is instructive to consider the lowest axial modes for the example of a resonator with a curvature of $\Delta k = 0.012~\mu$m$^{-1}$, a radius of $R_0 \approx 17.5~\mu$m, a resonant wavelength of around 850~nm, and an azimuthal quantum number of $m=180$, cf. Fig.~\ref{fig:tuning}~(a) for the axial intensity distribution. Increasing the axial quantum number causes the caustic position to shift to larger $z$ and thus $R_c$ decreases due to the resonator profile. At the same time, the resonance frequencies of the higher order modes shift to larger values.

\section{SPECTRAL TUNABILITY}
\label{sec:Tunability}
While equatorial WGMs have the advantage of having small mode volumes, they also exhibit a large frequency spacing between consecutive modes. In small WGM resonators, the azimuthal free spectral range is typically very large. For example, changing $m$ by one for a 35--$\mu$m diameter WGM changes its resonance frequency by $\Delta\nu_m\approx 1.9$~THz, i.e., about one percent of the optical frequency. Due to its monolithic design, tuning a WGM microresonator over such a large range is a critical issue. Electrical thermo-optic tuning of equatorial WGMs in a 75--$\mu$m diameter microtoroidal resonator over more than 300~GHz has been demonstrated for a wavelength of 1550~nm \cite{Armani04}. This corresponds to 35~\% of the azimuthal free spectral range (FSR) and to 0.15~\% of the optical frequency. Using a strain tuning technique, tuning over 400~GHz for a wavelength of 800~nm has been demonstrated for an 80--$\mu$m diameter microsphere, limited by the mechanical damage threshold of the resonator\cite{vonKlitzing01}. This corresponds to 50~\% of the azimuthal FSR and 0.12~\% of the optical frequency.

\begin{figure}
   \begin{center}
   \begin{tabular}{c}
   \includegraphics[height=7cm]{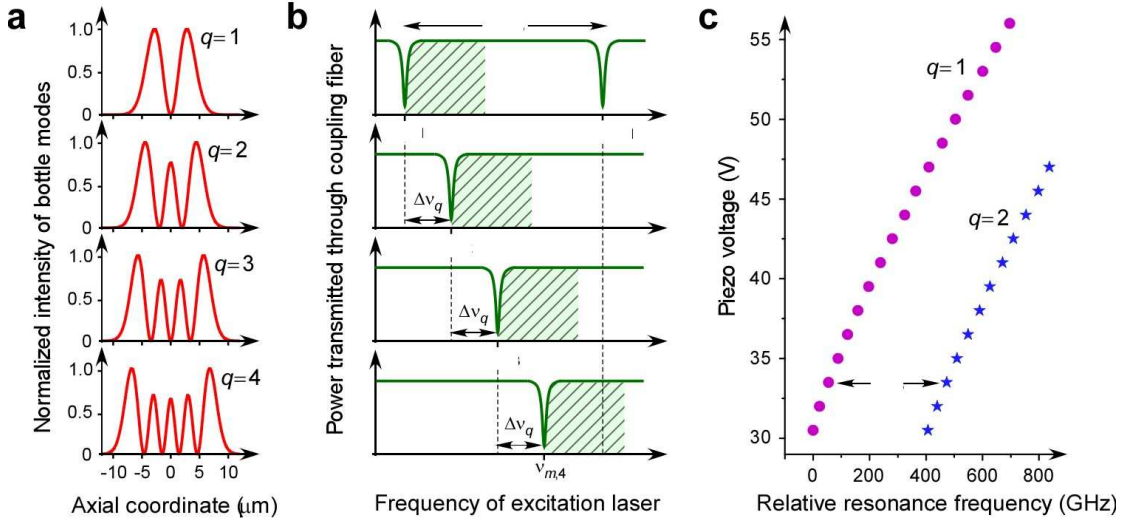}
   \end{tabular}
   \end{center}
   \caption[example]
   { \label{fig:tuning}
   Tuning the bottle microresonator. a) Calculated axial intensity distribution for TM polarized bottle modes with axial quantum numbers $q = 1, 2, 3, 4$ and azimuthal quantum number $m = 180$~. The calculated resonance wavelengths are 852.0~nm, 850.8~nm, 849.6~nm, and 848.3~nm, respectively. b)  Tuning scheme. Each of the bottle modes in (a) can be strain tuned by 700~GHz which exceeds the free spectral range (FSR) of $\Delta\nu_q =  \nu_{m,q+1}-\nu_{m,q} \approx c\Delta k/2\pi n = 397$~GHz between bottle modes of adjacent $q$ quantum numbers, where $c / n$ is the speed of light in the resonator material, and $n$ is the refractive index in vacuum. The resonance frequencies  $\nu_{m,q}$ and $\nu_{m+1,q}$ of modes differing by one in the azimuthal quantum number $m$ are spaced by $m = 1.92$~THz. This frequency interval can thus be bridged by four consecutive axial modes, making any arbitrary frequency accessible by using the set of modes $\{ \nu_{m,1}, \nu_{m,2}, \nu_{m,3}, \nu_{m,4}\}$ with $m$ properly chosen. c) Experimental demonstration of the tuning scheme in b) for the TM polarized $q = 1$ and $q = 2$ bottle modes. Piezo-electric shear actuators located on either end of the resonator fiber were used to elastically elongate the resonator. The strain tuning range of the $q = 1$ mode exceeds the $FSR$ $\Delta\nu_q$ by a factor of 1.6. The observed axial $FSR$ of $\Delta\nu_q = 425$~GHz is in good agreement with the theoretical value of 397~GHz.}
   \end{figure}

Recently, strain-tuning of a so-called ``microbubble resonator'' with a diameter of 200~$\mu$m over 690~GHz has been demonstrated \cite{Sum10Sup}. For $\lambda = 1550$~nm, this corresponds to 2.2 azimuthal FSRs. Microbubble resonators consist of a bulge on a microcapillary created from a silica tube\cite{Sum10Opt}. Due to the small wall thickness in the section forming the resonator, which is on the order of 1--2~$\mu$m, mechanical strain can be applied very efficiently to this structure, leading to the impressive frequency shift of 0.35~\% of the optical frequency. However, if one extrapolates these tuning results to resonator dimensions that may yield a good $Q/V$ ratio, such as 40~$\mu$m diameter, then the demonstrated resonator loses it full tunability. This is because the azimuthal FSR increases with smaller resonator dimensions. Moreover, the highest quality factor achieved in these structures is $1.5 \times 10^6$. Therefore, these resonators are not yet suitable for applications where high $Q/V$-ratios are required. Summarizing, full tunability of whispering-gallery-modes combining ultra-high quality factors and small mode volumes has so far not been achieved.

The bottle microresonator offers a solution for the problem described above. As mentioned, light harmonically oscillates in the bottle resonator between two turning points, resulting in a standing wave structure which can be compared to the one observed in Fabry--P\'erot microresonators where the light is reflected back and forth between two mirrors, cf. Fig.~\ref{fig_add_drop_configuration}~(b). This analogy goes even further: like the Fabry--P\'erot, the bottle microresonator exhibits an equidistant spectrum of eigenmodes, where neighboring modes differ only in the number of axial intensity nodes $q$. The corresponding spectral mode spacing $\Delta\nu_q$, called ``axial FSR'' in the following, is only dependent on the curvature of the resonator profile which can be customized during fabrication \cite{Poe09}. Also, like the Fabry--P\'erot, it is sufficient to tune the bottle microresonator over one axial FSR in order to have access to an arbitrary predetermined frequency with the appropriately chosen bottle mode. In our case, the axial FSR ($\sim0.4$~THz) is about a factor of four smaller than the azimuthal FSR of an equatorial WGM (typically less than 2~THz) of the same diameter.

In order to keep the mode volume small, it is desirable to work with a low axial quantum number $q$, thereby minimizing the separation between the two caustics. Remarkably, this is possible for the bottle microresonator which exhibits two FSRs, axial and azimuthal, and which can thus be tuned to any arbitrary frequency using, e.g., only the $q = 1$--4 axial bottle modes, cf. Fig.~\ref{fig:tuning}~(b). In Fig.~\ref{fig:tuning}~(c), tuning of a bottle mode over more than one axial FSR is presented in a 35--$\mu$m diameter resonator with a curvature of $\Delta k = 0.012~\mu$m$^{-1}$. The observed tuning range of 700~GHz is 1.6 times larger than the measured axial FSR of $\Delta\nu_q = (425 \pm 8)$~GHz. This corresponds to  a frequency shift of 0.2~\% of the optical frequency. The maximum strain applied to the resonator in this measurement, limited by the travel range of the tuning piezo, is only about 35~\% of the damage threshold of silica.

\begin{figure}
   \begin{center}
   \begin{tabular}{c}
   \includegraphics[width=0.5\textwidth]{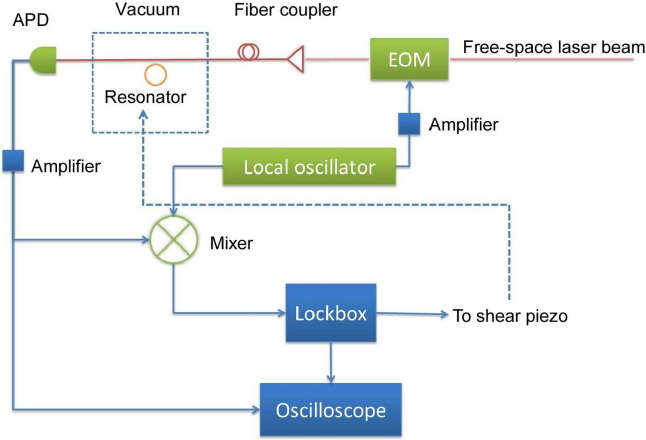}
   \end{tabular}
   \end{center}
   \caption[example]
   { \label{fig:freq_stab}
    Schematic of the Pound--Drever--Hall stabilization setup. APD: avalanche photodiode, EOM: electro-optic modulator.
    }
   \end{figure}
\section{FREQUENCY STABILIZED WGM MODES}
\label{sec:Stable}
Optical resonators are subject to environmental noise sources, such as temperature fluctuations and acoustic vibrations, that affect the short term as well as long term frequency stability. To counter this, active stabilization techniques have been developed, particularly in applications involving Fabry--P\'erot resonators. WGM resonators, despite their monolithic design, are also influenced by these noise sources and require active frequency stabilization. Two earlier works have demonstrated stabilization of WGM resonators via thermal self-stabilization and a lock-in technique, respectively \cite{Car04,Rez01}. These techniques have a number of drawbacks. Apart from the fact that the power is a constrained parameter using the self-stabilization technique, it may fluctuate and drift over time due to external perturbations, translating into frequency fluctuations and drifts of the resonator. In addition, it is not possible to directly lock to the center of the resonance. The lock-in technique is bandwidth-limited by the intensity build-up time of the resonator. Moreover, the recapture range, i.e., the frequency range over which the resonator remains locked while subject to disturbances, is determined by the resonator linewidth. To fulfill the need to stabilize a WGM resonator having ultra-high quality factor modes, we have developed a Pound--Drever--Hall locking scheme, cf. Fig.~\ref{fig:freq_stab}. It offers several advantages, namely, it effectively decouples power fluctuations from frequency fluctuations, has a large recapture range, and is not limited by the resonator bandwidth \cite{Dre83}.

An electro-optic modulator phase modulates the laser beam and is driven by a local oscillator running at 42.8~MHz. This frequency is large enough for the sidebands to be distinguishable from the carrier in the resonator spectrum, which has a linewidth of 2--4~MHz, dependent upon the particular resonator mode. Light, with as little as 15~nW of power, transmitted past the fiber-resonator coupling junction is detected on a high gain avalanche photodiode (APD). The signals from the local oscillator and APD are mixed, low-pass filtered, and sent to a proportional-integral (PI) controller where a correction signal is generated. This is in turn amplified and fed back to the resonator shear piezo. Care is taken not to excite mechanical resonances in the resonator fiber by using a 100~Hz low-pass filter before the piezo, thereby fixing the bandwidth of the control loop.

\begin{figure}
   \begin{center}
   \begin{tabular}{c c c}
   \includegraphics[width=0.25\textwidth]{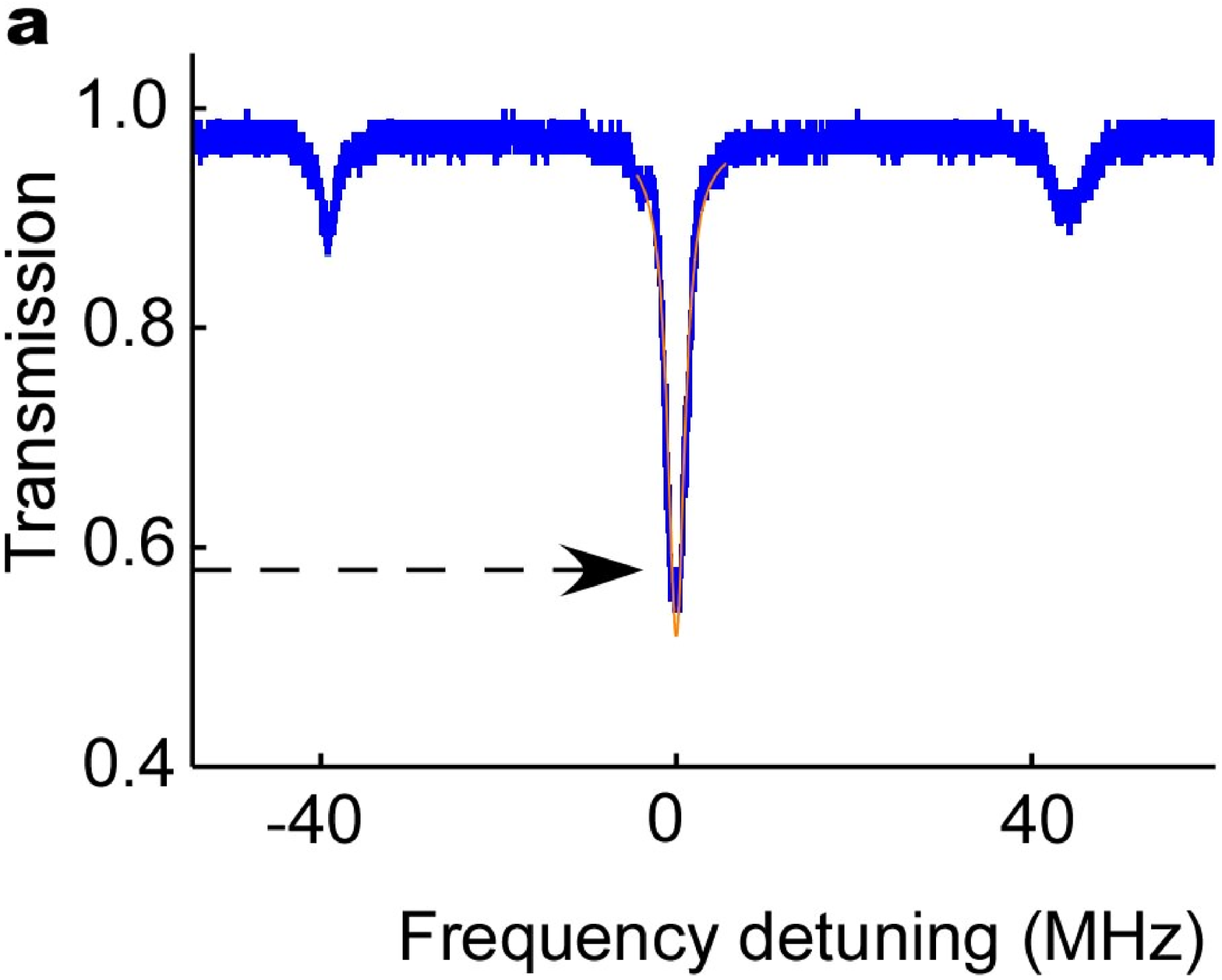} &
   \includegraphics[width=0.25\textwidth]{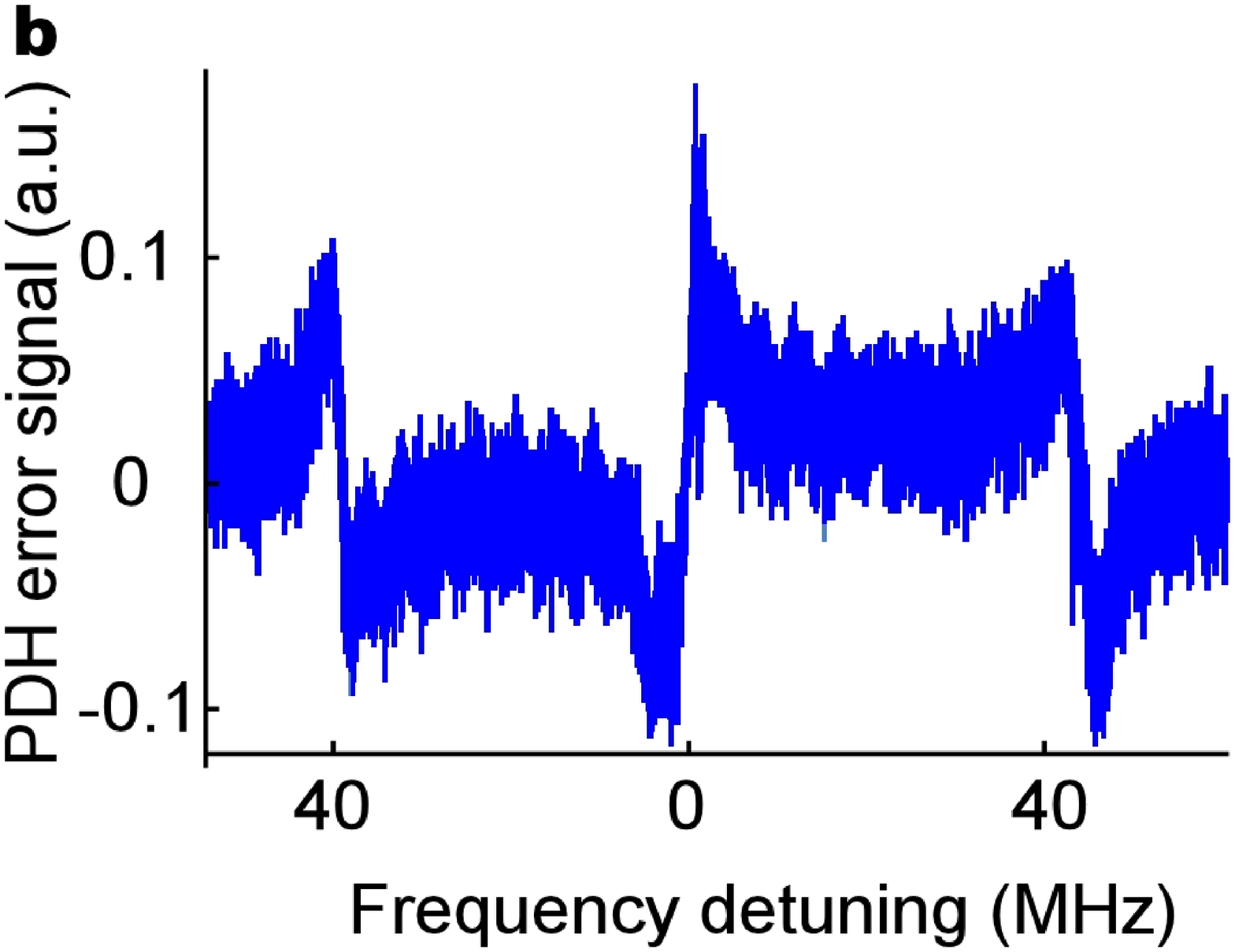} &
   \includegraphics[width=0.32\textwidth]{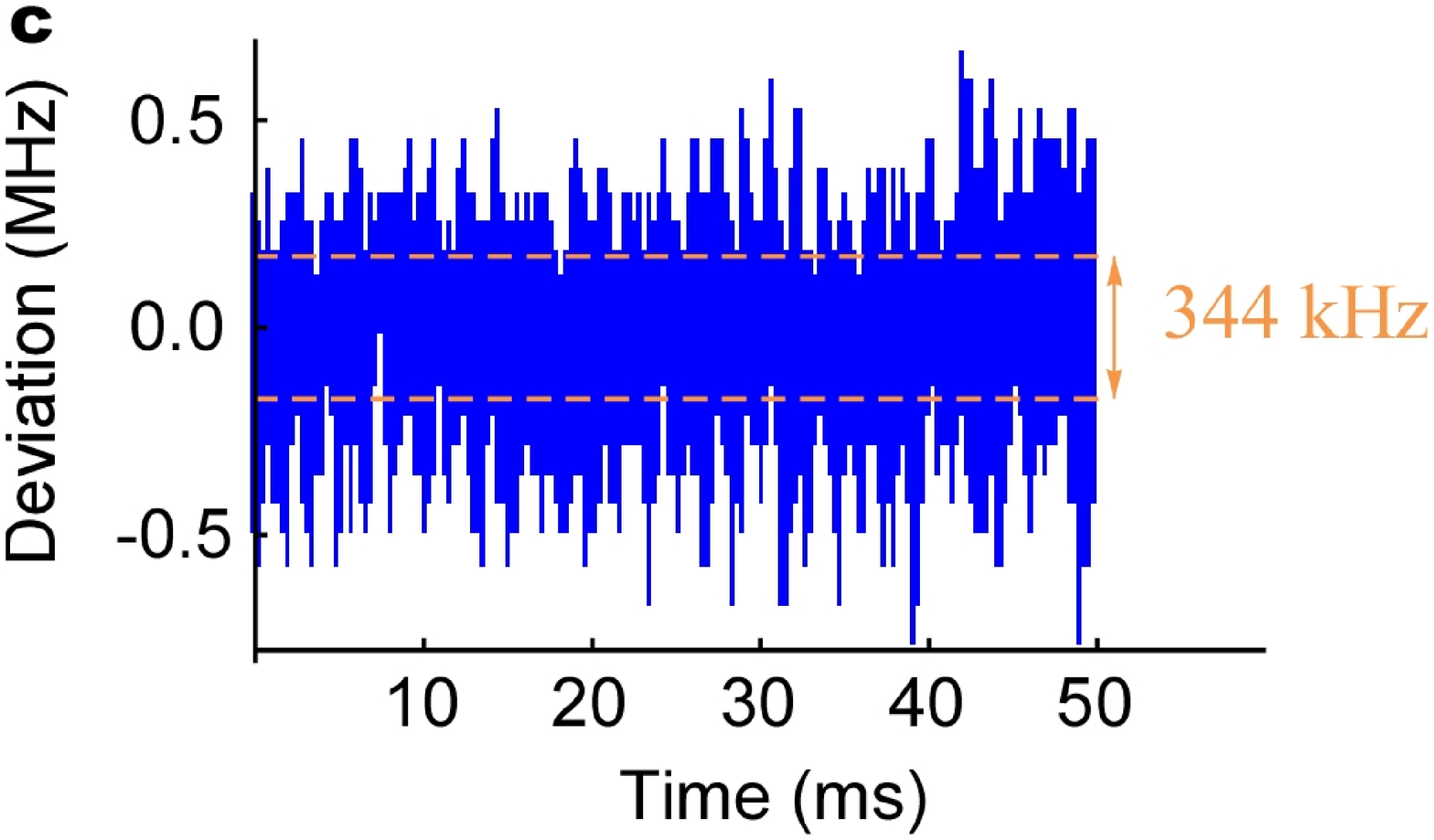}
   \end{tabular}
   \end{center}
   \caption[example]
   { \label{fig:freq_stab_data}
    Stabilization of an ultra-high $Q$ factor bottle mode in a vacuum of $2\times10^{-9}$~mBar. The mode lineshape (a) reveals a linewidth of 4.8~MHz ($Q_0=1.4\times10^8$) which is calibrated using sidebands as frequency markers. The corresponding PDH error signal is shown in (b). (c) When the resonator is stabilized to the laser, which is in turn stabilized to the D2 transition of $^{85}Rb$ using a Doppler-free spectroscopy setup, fluctuations are visible around the setpoint of the transmitted power. The rms value of these fluctuations is 344~kHz and is marked with dashed lines.
    }
   \end{figure}

We quantify the performance of the stabilization scheme in an ultra-high vacuum environment, compatible with the requirements of a single atom cavity quantum electrodynamics experiment. For this purpose, we measure the fluctuations of the transmitted power. Since these fluctuations are zero to first order when the resonator--laser detuning fluctuates around zero, we purposefully introduce an offset to the setpoint of the PI controller. This results in a non-zero resonator--laser detuning and relative frequency fluctuations between the laser and the resonator are thus directly converted into intensity fluctuations. The setpoint for the PI controller is indicated by the horizontal dashed line in Fig.~\ref{fig:freq_stab_data}~(a). With the resonator stabilized in Fig.~\ref{fig:freq_stab_data}~(c), fluctuations around the setpoint correspond to a rms frequency noise of 344~kHz or around 7~\% of the resonator linewidth over a time interval of 50~ms. This time interval is 5 times longer than the response time of the resonator lock. We have tested the lock for periods of up to one hour without any sign of degradation of its performance.

\subsection{Thermal Response in Vacuum}
Silica ranks among one of the lowest absorbing materials with as little as 0.2~dB/km loss at 1.5~$\mu$m wavelength. But, nevertheless, the absorbed light is converted into heat. This issue is particularly pertinent to monolithic microresonators where high $Q/V$ ratios facilitate huge power build-up factors and therefore extremely high power densities. For example, with just 1~$\mu$W of power coupled into a bottle resonator, the circulating intensity can reach 1~MW/cm$^2$. The heat dissipated in the mode volume influences the resonator radius, $R$, and refractive index, $n$, and this in turn shifts the resonance frequency, $\nu$, according to the relation
\begin{equation}
\frac{\delta\nu}{\nu}\approx-\frac{\Delta R}{R}-\frac{\Delta n}{n}.
\end{equation}
Placing the resonator in a vacuum exacerbates the issue by denying one of the heat dissipation routes, i.e., convection from the mode volume to the ambient environment. This leaves heat conduction through the resonator material as the dominant dissipation mechanism. While the resonator will reach a thermal equilibrium if simply probed with a CW laser light of constant power, many applications, such as the optical switch presented in section~\ref{sec:Switch}, involve pulsing or ramping the power. Accordingly, this causes the resonant frequency to change over time depending on the operating conditions.

\begin{figure}
   \begin{center}
   \begin{tabular}{c}
   \includegraphics[width=0.4\textwidth]{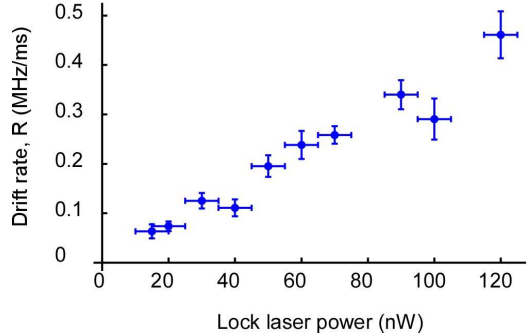}
   \end{tabular}
   \end{center}
   \caption[example]
   { \label{fig:drift}
    Dependence of the drift rate of a resonator mode on lock laser power. Increasing the power causes the absorption and therefore the drift rate to increase linearly.
    }
   \end{figure}

We have investigated the thermal response characteristics of a frequency stabilized bottle resonator having high quality factor and near critical coupling in vacuum. In particular, the step response of the resonator mode frequency is studied by first locking the resonator and then abruptly turning off the locking laser light. The development of the resonance frequency of the mode follows a typical exponential response according to the equation $\nu(t)=A\left(1-e^{-t/\tau}\right)$, where $A$ is a pre-factor in units of MHz and $\tau$ is the thermal time constant of the resonator. The setup used here is similar to that used for frequency locking the resonator in section~\ref{sec:Stable} except for the addition of a sample-and-hold circuit before the shear piezo that maintains the resonator at constant mechanical strain when the locking laser is turned off. An acousto-optic modulator performs the switching of the light with a rise/fall time of a few tens on nanoseconds. The drift of the resonance frequency of the mode is recorded by probing the transmission every 15~ms for a duration of 50~$\mu$s when the locking laser light is turned off. Simultaneously with the probing, the Pound--Drever--Hall error signal is recorded and reveals the direction in which the frequency of the mode has drifted. Figure~\ref{fig:drift} shows the drift rate, $R=A/\tau$, has a linear dependence with laser power and can be as much as 0.5~MHz/ms for a laser power of just 120~nW. For the resonator mode considered here, this power is enough to shift the resonance by about 10~\% of the linewidth in 2~ms. Extrapolating the results to higher powers indicates that with 1~$\mu$W of power the mode will be shifted by around 40~\% of the linewidth in 1~ms, thus affecting the transmission very strongly. A thermal time constant of $\tau=105$~ms was measured from data taken for powers of 15~nW and 20~nW. This is considerably longer than a similar measurement with a bottle resonator in air which found a time constant of 13--15~ms \cite{Poe10}. The difference is attributed to the relative importance of conductive and convective cooling mechanisms. For resonators with ultra high quality factor it is  thus desirable to use very low powers to lock the resonator. A feedforward mechanism to compensate the drift could be developed based on the presented data.

\section{ALL-OPTICAL SIGNAL PROCESSING USING THE KERR EFFECT}
\label{sec:Switch}
\begin{figure}
    \begin{center}
    \includegraphics[width=0.4\textwidth]{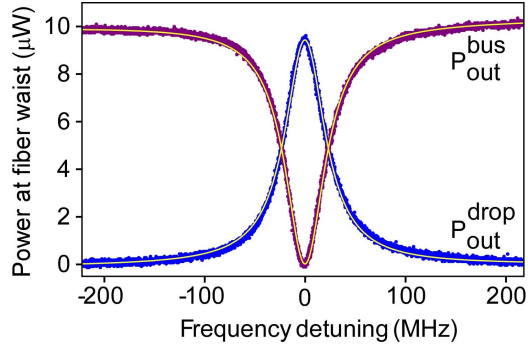}
    \end{center}
    \caption
    {\label{fig_power_transfer}
    Resonant power transfer between the two fibers coupled to the evanescent field of a bottle mode with ultra-high intrinsic quality factor. The plot shows the powers at the waists of the bus fiber $P_{\rm out}^{\rm bus}$ (purple dots) and the drop fiber $P_{\rm out}^{\rm drop}$ (blue dots) while the frequency of the probe laser is swept over the resonance. A loaded quality factor of $Q_{\rm load}=7.2 \times 10^6$ and a power transfer efficiency of 93~\% were inferred by fitting a Lorentzian to both signals (yellow curves).}
\end{figure}

One important application of the bottle microresonator lies in the field of all-optical signal processing where the flow of one light field is controlled using a second light field. As already mentioned in section~\ref{sec:intro}, the advantageous mode geometry of the bottle microresonator straight forwardly allows the realization of a so-called ``add--drop configuration'' where the resonator acts as a filter which selectively transfers light from a bus fiber to a drop fiber depending on the frequency of the light, cf. Fig.~\ref{fig_power_transfer}. All-optical switching can be implemented by using the third order nonlinearity of silica to control the resonance frequency of the resonator by changing the intra-cavity intensity.

In order to characterize the spectral properties and the transfer efficiency of the add--drop configuration on resonance, two ultra-thin fibers are placed at both caustics of a bottle mode with an intrinsic quality factor of $Q_{\rm 0}=1.8 \times 10^8$, cf. Fig.~\ref{fig_add_drop_configuration}. We investigate the system by means of a distributed feedback (DFB) diode laser which is scanned over the resonance of the bottle mode. The powers at both fiber output ports are monitored with photodiodes. As show in Fig.~\ref{fig_power_transfer}, when light is resonant with the frequency of the bottle mode, a Lorentzian-shaped dip in the detected power of the bus fiber occurs while light is simultaneously transferred to the drop fiber. Using a Lorentzian fit, we determine a linewidth of $\Delta\nu_{\rm load} = 49$~MHz, corresponding to a loaded quality factor of $Q_{\rm load}=7.2 \times 10^6$, and a transfer efficiency of 93~\% between the fiber waists. The overall transfer efficiency, including the losses at the taper transitions to the ultrathin fiber waists, remains as high as 90~\%. We note that the performance of our device in terms of combining high-efficiency filter functionality, high loaded quality factor, and single mode fiber operation lines up with the best to date \cite{Rok04Ult}.

\begin{figure}
    \begin{center}
    \includegraphics[width=0.5\textwidth]{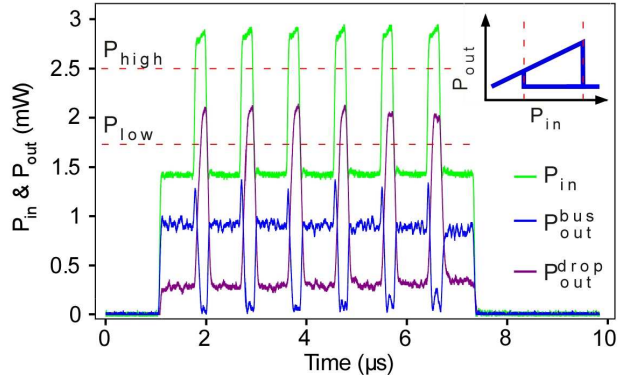}
    \end{center}
    \caption
    {\label{fig_switching}
    Demonstration of all-optical switching using the optical Kerr bistability in a  bottle microresonator. $P_{\rm in}$ (green) is varied between two levels which are located below and above the bistable regime (schematically indicated by the dashed lines) as shown in the inset. At the same time, the power at the outputs of the bus fiber (blue) and the drop fiber (purple) is monitored. As soon as the input power exceeds $P_{\rm high}$, i.e., the threshold power for optical bistability, 70~\% of the incident light is transferred to the output of the drop fiber. Lowering the input power below $P_{\rm low}$ again reverses the situation at the outputs.}
\end{figure}

Since the quality factor remains high in the add-drop configuration, extremely high intra-cavity intensities can be reached, even with very low input powers. The system is therefore an ideal candidate to demonstrate all optical switching schemes based on the nonlinear Kerr effect which connects the intra-cavity intensity with the refractive index. A variation of the intra-cavity intensity thus modifies the optical path length and thus changes the transmission properties of the bottle microresonator. In Fig.~\ref{fig_switching} we present routing of optical signals using this principle in the add--drop configuration by controlling the optical input power, $P_{\rm in}$. The laser frequency is initially detuned by -1.2 linewidths from the mode resonance. When $P_{\rm in}$ exceeds a threshold $P_{\rm high}$, the intracavity intensity reaches a critical value and the Kerr effect ``pulls'' the resonator mode into resonance with the laser frequency in a self-amplifying process. Thus, optical power is transferred from the bus to the drop fiber. Due to the nonlinearity of the bottle resonator material, the transmission properties exhibit a bistable behavior as indicated in the inset of Fig.~\ref{fig_switching}. In order to switch the signal back to the bus fiber the optical power therefore has to be decreased below another threshold value $P_{\rm low}$, well below $P_{\rm high}$, before the resonator returns to the initial situation. About 70~\% of the power launched into the bus fiber exits the drop fiber in the corresponding ``HIGH''-state with a modulation depth of 9~dB between the HIGH-state and the LOW-state while the switching rate can be as high as 1~MHz \cite{Poe10}.

To the best of our knowledge, this is the first time that a continuous wave signal is routed between two output channels using a single-wavelength scheme. All previous bistable single-wavelength schemes only demonstrated ON-OFF switching \cite{Pey85,He93}.

\section{SUMMARY}
Summarizing, we present a fully tunable WGM microresonator. The tunability of our bottle microresonator stems from the confinement of the light between two caustics in a simple Fabry--P\'erot-like axial mode structure. We have demonstrated active frequency stabilization of a bottle microresonator with an ultra-high quality factor in excess of $1\times10^8$ in an ultra-high vacuum environment using the Pound--Drever--Hall technique. Moreover, the advantageous mode geometry of the bottle modes allows near-lossless simultaneous coupling of two independent coupling fibers at the two caustics. We operated the resonator in this add-drop configuration and demonstrated a power transfer efficiency of up to 93~\% for a filter linewidth of only 49~MHz, corresponding to a loaded quality factor as high as $7.2\times10^6$. Finally, using the Kerr effect, we realized a bottle microresonator-based all-optical switch with an overall efficiency of 70~\% at rates of 1~MHz. In conjunction with its high $Q/V$ value, these applications reveal the enormous potential of the bottle microresonator for coupling light and matter. The bottle microresonator thus opens the route towards the realization of next-generation communication and information processing devices such as single photon all-optical switches\cite{Bermel06} and single photon transistors \cite{Chang07}.

\acknowledgments     

Financial support by the DFG (Research Unit 557), the Volkswagen Foundation, and the ESF (EURYI) is gratefully acknowledged.


\bibliography{report}   
\bibliographystyle{spiebib}   

\end{document}